\documentclass[11pt]{article}
\newtheorem{theorem}{Theorem}
\newtheorem{lemma}[theorem]{Lemma}

\newtheorem{corollary}[theorem]{Corollary}

\newtheorem{definition}[theorem]{Definition}

\usepackage{amsmath}
\usepackage{latexsym}
\usepackage{amsfonts}
\usepackage{amssymb}
\usepackage{setspace}
\usepackage{graphicx}

\newcommand{\R}{\mathbb{R}}

\newcommand{\Q}{\mathbb{Q}}

\newcommand{\qed}{\ensuremath{\Box}}
\newcommand{\up}[1]{\ensuremath{^\text{#1}}}
\newcommand\bpf[1][]{\smallskip{\bf Proof#1.}\quad}
\newcommand\epf{\qed\medskip}

\newcommand\nnz{\hbox{\sf nnz}}

\newcommand\opt{\hbox{opt}}
\newcommand\argmin{\hbox{argmin}}
\newcommand\supp{\hbox{supp}}
\newcommand\rnk{\hbox{\sf rank}}

\newcommand\col{\hbox{\sf col}}
\newcommand\row{\hbox{\sf row}}
\newcommand\sgn{\hbox{\sf sgn}}

\newcommand\nll{\hbox{\sf null}} 

\newcommand\etal{{\it et al.\ }}

\newcommand{\problem}[2]{\par\noindent\hbox{\parbox{390pt}{{\sf\large #1}\quad #2}}\par\smallskip}

\begin{document}

\title{Matrix sparsification and the sparse null space problem}
\author{\em Lee-Ad Gottlieb
\thanks{Weizmann Institute of Science, Rehovot, Israel.
E-mail: {\tt lee-ad.gottlieb@weizmann.ac.il}.}
\and \em Tyler Neylon
\thanks{Bynomial, Inc.
E-mail: {\tt tylerneylon@gmail.com}.}}


\maketitle

\begin{abstract} \noindent
We revisit the matrix problems {\sf sparse null space} and
{\sf matrix sparsification}, and show that they are equivalent.
We then proceed to seek algorithms for these problems: We prove the hardness
of approximation of these problems, and also give a powerful tool to
extend algorithms and heuristics for sparse approximation theory to these
problems.
\end{abstract}


\section{Introduction}

In this paper, we revisit the matrix problems {\sf sparse null space} and
{\sf matrix sparsification}.

The {\sf sparse null space} problem was first considered by Pothen in 1984
\cite{P-84}. The problem asks, given a matrix $A$, to find a matrix $N$
that is a full null matrix for $A$ -- that is, $N$ is full rank and the
columns of $N$ span the null space of $A$. Further, $N$ should be sparse,
i.e. contain as few nonzero values as possible. 
The {\sf sparse null space} problem is motivated by its use
to solve Linear Equality Problems (LEPs) \cite{CP-86}. LEPs arise in the
solution of constrained optimization problems via generalized gradient
descent, segmented Lagrangian, and projected Lagrangian methods.
Berry \etal \cite{BH-85} consider the {\sf sparse null space} problem in the
context of the dual variable method for the Navier-Stokes equations, or more
generally in the context of null space methods for quadratic programming.
Gilbert and Heath \cite{GH-87} noted that among the numerous applications
of the {\sf sparse null space} problem arising in solutions of underdetermined
system of linear equations, is the efficient solution to the force method (or
flexibility method) for structural analysis, which uses the null space to
create
multiple linear systems. Finding a sparse null space will decrease the run
time and memory required for solving these systems.
More recently, it was shown \cite{ZZ-05,N-06} that the {\sf sparse null
space}
problem can be used to find correlations between small numbers of times
series,
such as financial stocks.
The decision version of the {\sf sparse null space} problem is known to be 
NP-Complete
\cite{CP-86}, and
only heuristic solutions have been suggested for the minimization problem 
\cite{CP-86,GH-87,BH-85}.

The {\sf matrix sparsification} problem is of the same flavor as
{\sf sparse null space}. One is given a full rank matrix $A$, and the task
is to find another matrix $B$ that is equivalent to $A$ under elementary
column operations, and contains as few nonzero values as possible.
Many fundamental matrix operations are greatly simplified by first
sparsifying
a matrix (see \cite{DER-86}) and the problem has applications in areas
such as machine learning \cite{SS-00} and in discovering cycle bases of
graphs \cite{KMMP-04}.
But there seem to be only a small number of heuristics
for {\sf matrix sparsification} (\cite{CM-92} for example), or algorithms
under
limiting assumptions (\cite{H-84} considers matrices that satisfy the Haar
condition),
but no general approximation algorithms. McCormick \cite{M-83} established
that the decision version of this
problem is NP-Complete.

For these two classic problems, we wish to investigate potentials and limits
of approximation algorithms both for the general problems and for some
variants under simplifying assumptions. To this end, we will need to consider the
well-known
vector problems {\sf min unsatisfy}
and {\sf exact dictionary representation} (elsewhere called the
{\sf sparse approximation} or {\sf highly nonlinear approximation}
problem \cite{T-04}).

The {\sf min unsatisfy} problem is an intuitive problem on linear
equations. Given a system $Ax=b$ of linear equations (where $A$ is an integer
$m \times n$ matrix and $b$ is an integer $m$-vector), the problem is to
provide a
rational $n$-vector $x$; the measure to be minimized is the number of
equations
not satisfied by $Ax=b$. The term ``min unsatisfy'' was first coined by
Arora \etal
\cite{ABSS-97} in a seminal paper on the hardness of approximation, but
they claim
that the the NP-Completeness of the decision version of this problem is 
implicit in a 1978 paper of
Johnson and Preparata \cite{JP-78}. Arora \etal demonstrated that it is 
hard
to approximate {\sf min unsatisfy} to within a factor $2^{\log^{.5-o(1)} n}$ of 
optimal (under the assumption that NP does not admit a quasi-polynomial time
deterministic algorithm).
This hardness result holds over $\mathbb{Q}$, and stronger results are known for
finite fields \cite{DKRS-03}.
For this problem, Berman and Karpinski \cite{BK-01} gave a randomized
$\frac{m}{c \log m}$-approximation algorithm (where $c$ is a constant). We
know of no heuristics studied for this problem.

The {\sf exact dictionary representation} problem is the fundamental problem
in sparse approximation theory (see \cite{M-04}). In this problem, we are
given a
matrix of dictionary vectors $D$ and a target vector $s$, and the task is
to find
the smallest set
$D' \subset D$ such that a linear combination of the vectors of $D'$ is
equal to $s$.
This problem and its variants have been well studied. According to
Temlyakov \cite{T-03}, a
variant of this problem may be found as early as 1907, in a paper of
Schmidt \cite{S-07}.
The decision version of this problem was shown to be NP-Complete by 
Natarajan \cite{N-95}. (See
\cite{MG-07} for further discussion.)

The field of sparse approximation theory has become exceedingly
popular:
For example, SPAR05 was largely devoted to it, as was the SparseLand 2006
workshop
at Princeton, and a mini-symposium at NYU's Courant Institute in
2007. The applications
of sparse approximation theory include signal representation
and recovery
\cite{CW-92,NT-09},
amplitude optimization \cite{S-89} and function approximation \cite{N-95}.
When the dictionary vectors are Fourier coefficients, this problem is
a classic problem
in Fourier analysis, with applications in data compression, feature
extraction, locating approximate periods and similar data mining problems
\cite{ZGSD-06,GGIMS-02,GMS-04,CRT-06}. There is a host of results for
this problem,
though all are heuristics or approximations under some qualifying
assumptions.
In fact, Amaldi and Kann \cite{AK-95} showed that this problem (they
called it
{\sf RVLS} -- `relevant variables in the linear system') is as hard to
approximate
as {\sf min unsatisfy}, though their result seems to have escaped the
notice of
the sparse approximation theory community.

\noindent{\bf Our contribution.} As a first step, we note that the
matrix problems
{\sf sparse null space} and {\sf matrix sparsification} are equivalent,
and that the vector
problems {\sf min unsatisfy} and {\sf exact dictionary representation} are
equivalent as
well. 
Note that although these equivalences are straightforward, they 
seem to have escaped researchers in this field. For example, 
\cite{BFP-95} claimed that the
{\sf sparse null space} problem is computationally more difficult than
{\sf matrix sparsification}.)

We then  
proceed to show that {\sf matrix sparsification} is hard to
approximate, via a reduction from {\sf min unsatisfy}. We will thereby 
show that the two matrix problems are hard to approximate within a 
factor $2^{\log^{.5-o(1)} n}$ of optimal (assuming NP does not admit 
quasi-polynomial time deterministic algorithms).

This hardness result for {\sf matrix sparsification} is important in its
own right, but it
further leads us to ask what {\em can} be done for this problem.
Specifically, what restrictions
or simplifying assumptions may be made upon the input matrix to make {\sf
matrix sparsification}
problem tractable? In addressing this question, we provide the major
contribution of this paper and show how to adapt the vast number of
heuristics and algorithms for {\sf exact dictionary representation} to
solve {\sf matrix sparsification} (and hence {\sf sparse null space} as well).
This allows us to conclude, for
example, that {\sf matrix sparsification} admits a randomized $\frac{m}{c
\log m}$-approximation algorithm, and also to give limiting conditions
under which a known $\ell_1$ relaxation scheme for
{\sf exact dictionary matching} solves {\sf matrix sparsification} exactly.
Our results also carry over to relaxed version of these problems, where
the input is extended by an error term $\delta$ which relaxes a
constraint.  These versions are defined in the appendix (\S\ref{relaxed_section}),
although we omit the proof of their equivalence. 
All of our results assume that the vector variables are over $\mathbb{Q}$.

An outline of our paper follows: In Section \ref{pre} we review some 
linear algebra and introduce notation. 
In closing the preliminary 
section (\ref{equiv}), we prove equivalences between the two 
matrix problems and the two vector problems. 
In Section~\ref{matrix-hard} 
we prove that {\sf matrix sparsification} is hard to approximate, and in 
Section~\ref{relation} we show how to adapt algorithms for {\sf exact 
dictionary representation} to solve {\sf matrix sparsification}.

\section{Preliminaries}\label{pre}

In this section we review some linear algebra, introduce notation and 
definitions, and formally state our four problems. 

\subsection{Linear algebra and notation.}


\noindent{\bf Matrix and vector properties.}
Given a set $V$ of $n$ $m$-dimensional column vectors, an $m$-vector
$v \notin V$ is {\em independent} of the vectors of $V$ if there is no
linear combination of
vectors in $V$ that equals $v$. A set of vectors is independent if each
vector in
the set is independent of the rest.

Now let the vectors of $V$ be arranged as columns of an $m \times n$
matrix $A$;
we refer to a column of $A$ as $a_i$, and to a position in $A$ as $a_{ij}$.
We define \#col($A$) to be the number of columns of $A$.
The column {\em span} of $A$ (col($A$)) is the (infinite) set of column
vectors
that can be produced by a linear combination of the columns of $A$.
The {\em column rank} of $A$ is the dimension of the column space of $A$
(rank($A$) $=$ dim(col($A$))); it is the size of the maximal independent
subset in
the columns of $A$. If the column rank of $A$ is equal to $n$, then the
columns of $A$ are independent, and $A$ is said to be {\em full rank}.

Other matrices may be produced from $A$ using {\em elementary column
operations}. These include multiplying columns by a nonzero factor,
interchanging columns, and adding a multiple of one column to another. These
operations produce a matrix $A'$ which has the same column span as $A$;
we say $A$ and $A'$ are {\em column equivalent}. It
can be
shown that $A$, $A'$ are column equivalent iff $A'=AX$ for some invertible
matrix $X$.

Let $R$ be a set of rows of $A$, and $C$ be a set of columns. $A(R,C)$ is
the submatrix of $A$
restricted to $R$ and $C$. Let $A(:,C)$ ($A(R,:)$) be the submatrix of $A$
restricted to all rows
of $A$ and to columns in $C$ (restricted to the rows of $R$ and all
columns in $A$).
A {\em square matrix} is an $m \times m$ matrix. A square matrix is {\em
nonsingular} if it is invertible.

\noindent{\bf Null space.}
The {\em null space} (or {\em kernel}) of $A$ (null(A)) is the set of all
nonzero
$n$-length vectors $b$ for which $Ab = 0$. The rank of $A$'s null space is
called
the {\em corank} of $A$. The rank-nullity theorem states that for any
matrix $A$,
rank$(A)+$ corank$(A) = n$. Let $N$ be a matrix consisting of column
vectors in
the null space of $A$; we have that $AN = 0$. If the rank of $N$ is equal
to the
corank of $A$ then $N$ is a {\em full null matrix} for $A$.

Given matrix $A$, a full null matrix for $A$ can be constructed in
polynomial time. Similarly,
given a full rank matrix $N$, polynomial time is required to construct a
matrix $A$ for which
$N$ is a full null matrix \cite{N-06}.

\noindent{\bf Notation.}
Throughout this paper, we will be interested in the number of zero and
nonzero
entries in a matrix $A$. Let \nnz(A) denote the number of nonzero entries
in $A$.
For a vector $x$, let $||x||_0$ denote the number of nonzero entries in
$x$. This
notation refers to the quasi-norm $\ell_0$, which is not a true norm since
$\lambda||x||_0 \ne ||\lambda x||_0$, although it does honor the triangle inequality.

For vector $x$, let $x_i$ be the value of the $i$\up{th} position in $x$.
The {\em support}
of $x$ (\supp(x)) is the set of indices in $x$ which correspond to nonzero
values,
$i \in supp(x) \Leftrightarrow x_i \ne 0$.

The notation $A|B$ indicates that the rows of matrix $B$ are concatenated to 
the rows of matrix $A$. The notation $\binom{A}{B}$ indicates that the 
columns of $B$ are appended to the columns of $A$. $M=A \otimes B$ denotes 
the Kronecker product of two matrices, where $M$ is formed by multiplying 
each individual entry in $A$ by the entire matrix $B$. (If $A$ is $m\times 
n$, $B$ is $p\times q$, then $M$ is $mp\times nq$.)

By {\em equivalent} problems, we mean that reductions between them 
preserve approximation factors. A formal definition of approximation 
equivalence is 
found in Appendix \ref{app:equiv}. 

\subsection{Minimization problems}

In this section, we formally state the four major minimization problems 
discussed in this paper. The first two problems have vector solutions, and 
the second two problems have matrix solutions. Our results hold when the 
variables are over $\mathbb{Q}$, although these problems can be defined 
over $\R$. $\mathcal{I}_F$ is the set of input instances, $S_F(x)$ is the 
solution space for $x \in \mathcal{I}_F$, $M_F(x,y)$ is the objective 
metric for $x \in \mathcal{I}_F$ and $y \in S_F(x)$.

\problem{exact dictionary representation (EDR)}{\\
$\mathcal{I}_\mathsf{EDR} = \langle D,s\rangle$,
$m\times n$ matrix $D$, vector $s$ with $s\in \col(D)$\\
$S_\mathsf{EDR}(D,s) = \{v\in \Q^n : Dv = s \}$ \\
$m_\mathsf{EDR}(\langle D,s\rangle, v) = ||v||_0$
}


\problem{min unsatisfy (MU)}{\\
$\mathcal{I}_\mathsf{MU} = \langle A,y\rangle$,
$m\times n$ matrix $A$, vector $y\in \Q^m$\\
$S_\mathsf{MU}(A,y) = \{x:x\in \Q^n \}$ \\
$m_\mathsf{MU}(\langle A,y\rangle, x) = ||y-Ax||_0$
}

\problem{sparse null space (SNS)}{\\
$\mathcal{I}_\mathsf{SNS}  = $ matrix $A$\\
$S_\mathsf{SNS}(A) = \{ N : N$ is a full null matrix for $ A\}$ \\
$m_\mathsf{SNS}(A,N) = \nnz(N)$}

\problem{matrix sparsification (MS)}{\\
$\mathcal{I}_\mathsf{MS} = $ full rank $m\times n$ matrix $B$ \\
$S_\mathsf{MS}(B) = \{ $matrix $N: N=BX$ for some invertible matrix $ X 
\}$ \\
$m_\mathsf{MS}(B,N) = \nnz(N)$
}




\subsection{Equivalences}\label{equiv}

In closing the preliminary section, we show that {\sf min unsatisfy}
and {\sf exact dictionary representation} are equivalent. 
We then show that {\sf matrix sparsification} and
{\sf matrix sparsification} are equivalent.
The type of equivalence is formally stated in definition \ref{equiv_def}, and
guarantees exact equality of approximation factors among polynomial-time
algorithms (Corollary \ref{equiv_cor}).

\paragraph{Equivalence of vector problems.}

Here we show that {\sf EDR} and {\sf min unsatisfy} are
equivalent.

We reduce {\sf EDR} to {\sf min unsatisfy}. Given input $\langle
D,s\rangle$ to
{\sf EDR}, we seek a vector $v$ with minimum $||v||_0$ that
satisfies
$Dv = s$. Let $y$ be any vector that satisfies $Dy=s$, and $A$ be a full null
matrix for $D$. (These can be derived in polynomial time.) Let $x =
MU(A,y)$ and
$v = y-Ax$. We claim that $v$ is a solution to {\sf EDR}. First note that $v$
satisfies $Dv=s$: $Dv = D(y-Ax) = Dy-DAx = s-0 = s$. Now,
the call
to $MU(A,y)$ returned a vector $x$ for which $||y-Ax||_0 = ||v||_0$ is
the minimization measure; and, as $x$ ranges over $\R^n$, the vector
$v=y-Ax$ ranges
over all vectors with $Dv=s$.  Hence, the oracle for {\sf min unsatisfy}
directly
minimizes $||v||_0$, and so $v$ is a solution to {\sf EDR}.




We now reduce {\sf min unsatisfy} to {\sf EDR}. Given input $\langle A,y
\rangle$ to
{\sf min unsatisfy}, we seek a vector $x$ which minimizes
$||y-Ax||_0$.
We may assume that $A$ is full rank. (Otherwise, we can simply take
any matrix $\tilde{A}$ whose columns form a basis of $\col(A)$, and it
follows easily that
$||MU(A,y)||=||MU(\tilde{A},y)||$.) Find (in polynomial time) a matrix $D$
such
that $A$ is a full null matrix for $D$ (this can be achieved by finding
$D^T$ as a null matrix of $A^T$). Let $s=Dy$, and
$v=\mathsf{EDR}(D,s)$.
Since $Dv=s$ we have that $D(y-v) = Dy-Dv =0$, from which we
conclude that
$y-v$ is in the null space of $D$, and therefore in the column space of
$A$. It
follows that we can find an $x$ such that $Ax=y-v$. We claim that $x$
solves the
instance of {\sf min unsatisfy}: It suffices to note that the call to
$\mathsf{EDR}(D,s)$ minimizes
$||v||_0 = ||y-Ax||_0$,
and that as $v$ ranges over $\{v:Dv=s\}$, the vector
$Ax = y-v$ ranges over all of $\col(A)$.
In conclusion,

\begin{lemma}
The problems {\sf exact dictionary representation} and {\sf min unsatisfy} 
are equivalent.
\end{lemma}

\paragraph{Equivalence of matrix problems.}

Here we demonstrate that
{\sf sparse null space} and {\sf matrix sparsification} are equivalent.
Recall that in the description of {\sf matrix sparsification} on
input matrix $B$, we required that $B$ be full rank,
$\#\col(B) = \rnk(B)$. (We could in fact allow $\#\col(B) > \rnk(B)$,
but this would trivially result in $\#\col(B) - \rnk(B)$ zero columns in
the solution, and these columns are not interesting.) We will need the
following lemma:

\begin{lemma}\label{ns-lemma}
Let $B$ be a full null matrix for $m \times n$ matrix $A$.
The following statements are equivalent:
(1) $N=BX$ for some invertible matrix $X$.
(2) $N$ is a full null matrix for $A$.
\end{lemma}

\bpf
In both cases, $N$ and $B$ must have the same number of columns,
the same rank, and the same span.  This is all that is required to demonstrate
either direction.
\epf

We can now prove that  {\sf sparse null space} and {\sf matrix
sparsification}
are equivalent. The problem {\sf sparse null space} may be solved
utilizing an oracle for {\sf matrix sparsification}. Given input $A$ to
{\sf sparse null space}, create (in polynomial time) a matrix $B$ which is a
full null matrix for $A$, and let $N = \mathsf{MS}(B)$. We claim that $N$
is a solution
to $\mathsf{SNS}(A)$. Since $N=BX$ for some invertible matrix $X$, by
Lemma~\ref{ns-lemma} $N$ is a full null matrix for $A$.
Therefore the call to $\mathsf{MS}(B)$ is equivalent to a call
to $\mathsf{MS}(N)$, which solves {\sf sparse null space} on $A$.

We show that {\sf matrix sparsification} can be solved using an oracle
for {\sf sparse null space}. Given input $B$ to {\sf matrix sparsification},
create (in polynomial time) matrix $A$ such that $B$ is a full null matrix
for $A$. Let $N = \mathsf{SNS}(A)$.  We claim that $N$ is a solution to
$\mathsf{MS}(B)$. By the lemma, $N=BX$ for some invertible matrix $X$, so $N$
can be derived from $B$ via elementary row reductions.
The call to $\mathsf{SNS}(A)$ finds an optimally sparse $N$, which is
equivalent to solving {\sf min unsatisfy} on $B$.
In conclusion,

\begin{lemma}
The problems
{\sf matrix sparsification} and {\sf sparse null space} are equivalent.
\end{lemma}

\section{Hardness of approximation for matrix problems}\label{matrix-hard}

In this section, we prove the hardness of approximation of {\sf matrix
sparsification} (and therefore {\sf sparse null space}). This motivates
the search for heuristics or algorithms under simplifying assumptions for
{\sf matrix sparsification}, which we undertake in the next section. For the 
reduction, we will need a relatively
dense matrix which we know cannot be further sparsified. We will prove the
existence of such a matrix in the first subsection.  

\subsection{Unsparsifiable matrices}

Any $m\times n$ matrix $A$ may be column reduced to contain at most
$(m-r+1)r$ nonzeros, where $r=\rnk(A)$. For example, Gaussian elimination
on the columns of the matrix will accomplish this sparsification.
We will say that a rank $r$, $m\times n$ matrix $A$ is
{\em completely unsparsifiable} if and only if, for any invertible matrix
$X$,
$\nnz(AX) \ge (m-r+1)r$. A matrix $A$ is {\em optimally sparse} if, for any
invertible $X$, $\nnz(AX) \ge \nnz(A)$. The main result of this section
follows.

\begin{theorem}\label{unsparsifiable_theorem}
Let $A$ be an $m\times n$ matrix with $m \ge n$. If every square submatrix
of $A$ is
nonsingular, then $A$ has rank $n$ and is completely unsparsifiable. 
Moreover, in such
case the matrix $\binom{I}{A}$ is optimally sparse, where $I$ is the
$n\times n$ identity
matrix.
\end{theorem}


Before attempting a proof of the theorem, we need a few intermediate results.

\begin{lemma}\label{opt_sparse_lemma}
Matrix $A$ is optimally sparse if and only if, for any vector $x\ne 0$,
$||Ax||_0 \ge \max_{i\in\supp(x)} ||a_i||_0.$
\end{lemma}

\bpf
Suppose that there exists an $x$ that, for some $i\in\supp(x)$,
$||Ax||_0 < ||a_i||_0$. Then we may replace the matrix column
$a_i$ by $Ax$, and create a matrix with the same rank as $A$ which
is sparser than $A$; a contradiction. Similarly, suppose that $A$
is not optimally sparse, so that there exists $B=AX$ with $\nnz(B) <
\nnz(A)$,
for some invertible $X$. Assume without loss of generality that the
diagonal of $X$ is full, $x_{ii}\ne 0$ (otherwise just permute the
columns of $X$ to make it so).  Then there must exist an index $j\in [n]$
with $||b_j||_0 < ||a_j||_0$, and we have
$||Ax_j||_0 = ||b_j||_0 < ||a_j||_0 \le \max_{i\in\supp(x_j)}||a_i||_0$,
since $x_{jj} \ne 0$.
\epf


A submatrix $A(R,C)$ is {\em row-inclusive} iff
$r\not\in R$ implies that $A(r,C)$ is not in the row span of $A(R,C)$.
In other words, $A(R,C)$ includes all the rows of $A(:,C)$ which are
in the row span of this submatrix.
A submatrix $A(R,C)$ is a {\em candidate submatrix} of $A$ (written $A(R,C)\lhd
A$) if
and only if $A(R,C)$ is both row-inclusive and $\rnk(A(R,C)) = |C|-1$.
This last property is equivalent to stating that the columns of $A(R,C)$
form a {\em circuit} -- they are minimally linearly dependent.
We can potentially zero out $|R|$ entries of $A$ by using the column dependency
of $A(R,C)$; being row-inclusive means there would be exactly $|R|$
zeros in the modified column of $A$.

The next lemma demonstrates the close relationship between candidate
submatrices
and vectors $x$ which may sparsify $A$ as in Lemma \ref{opt_sparse_lemma}.

\begin{lemma}\label{cand_submatrix_lemma}
For any $m\times n$ matrix $A$:
(1) For any $x\ne 0$ and $i\in\supp(x)$, there exists $A(R,C)\lhd A$ for which
$|R| \ge m - ||Ax||_0$, and $i\in C\subset\supp(x)$.
(2) For any $A(R,C)\lhd A$ there exists a vector $x$ for which
$\supp(x) = C$ and $||Ax||_0 = m - |R|$.
\end{lemma}

\bpf
Part 1: Let $R' = [m] - \supp(Ax)$ (where [m] = \{1, 2, ..., m\}), 
and choose $C$ so that $i\in
C\subset\supp(x)$,
and the columns of $A(R',C)$ form a circuit. (Note that the columns
$A(R',\supp(x))$ are
dependent since $A(R',:)x=0$).  Now expand $R'$ to $R$ so that
$A(R,C)$ is row-inclusive.  Then $\rnk(A(R,C)) = \rnk(A(R',C)) = |C|-1$,
so that
$A(R,C)\lhd A$.

Part 2: Since the columns of $A(R,C)$ form a circuit, there is an $\tilde
x$ with $\tilde x_i\neq 0\,\forall i$ and
$A(R,C)\tilde x = 0$.  Then
$\dim(\col(A(R,C)^T)) = |C| - 1 = \dim(\nll(\tilde x^T))$
and also $\col(A(R,C)^T) \subset \nll(\tilde x^T)$, which together
imply $\col(A(R,C)^T) = \nll(\tilde x^T)$.
So $A(r,C)\tilde x = 0$ is true iff $r\in R$ (using the fact that
$A(R,C)$ is
row-inclusive). Now choose $x$ so that $x(C) = \tilde x$ and all other
coordinates
are zero; then $\supp(Ax) $ $= [m] - R$.
\epf

The following is an immediate consequence of the lemma, and is crucial to
our proof
of Theorem~\ref{unsparsifiable_theorem}.

\begin{corollary}\label{final_optimally_sparse_lemma}
The $m\times n$ matrix $A$ is optimally sparse if and only if there is no
candidate submatrix $A(R,C)\lhd A$ with $m-|R| < ||a_i||_0$ for
some $i\in C$.
\end{corollary}

We are now ready to prove the theorem.

\bpf[ of Theorem \ref{unsparsifiable_theorem}]
Let $B = \binom{I}{A}$.  We prove that $B$ is optimally sparse.
Suppose $B(R,C)\lhd B$.  Let $R_I = R\cap [n]$ and $R_A = R-[n]$.
Now $B(R_I,C)$ is a submatrix of $I$ with dependent columns, so
$B(R_I,C) = 0$.  By row-inclusiveness, $R_I$ must include all zero
rows in $B([n],C)$, so $|R_I| = n - |C|$.
Since $B(R_I,C) = 0$, it follows that $\rnk(B(R_A,C)) = \rnk(B(R,C)) =
|C|-1$, and
$|R_A| \ge {|C|} - 1$.  Any $|C|\times |C|$ subsquare of $B(R_A,C)$ would make
the rank at least $|C|$, so we must have $|R_A| < |C|$; thus $|R_A| = |C|-1$.
Combined with $|R_I| = n - |C|$, this implies that $|R| = n - 1$. Then
$m+n-|R| = m+1 = ||b_i||_0$ for any column $b_i$ of $B$, proving that $B$
is optimally sparse by corollary \ref{final_optimally_sparse_lemma}.

Recall that Gaussian elimination on matrix $A\to G$ yields
$\nnz(G) = (m-n+1)n$.  Now suppose there is an invertible matrix $X$
with $\nnz(AX) < (m-n+1)n$.  Then $\nnz(BX) = \nnz(\binom{X}{AX})
< n^2 + (m-n+1)n = (m+1)n$, contradicting the optimal sparsity of $B$.
Hence no such $X$ exists and $A$ is completely
unsparsifiable.
\epf

\subsection{Efficiently building an unsparsifiable matrix}

The next lemma establishes that we can easily construct
an unsparsifiable matrix with a given column, a useful fact
for the reductions to follow.

\begin{lemma}\label{lem:const_opt}
If $n\times n$ matrix $M = (M_{ij})$ has entries $m_{ij} = i ^ {p_j}$
for distinct positive reals $p_1, p_2, \ldots, p_n$, then every subsquare of
$M$ is nonsingular.
\end{lemma}

\bpf
Let $f$ be a {\em signomial} (a polynomial allowed to have nonintegral
exponents).  We define 
$\mathsf{positive\_zeros}(f) := \{x : x > 0 \;\&\; f(x) = 0\}$
and $\#\mathsf{sign\_changes}(f) := \#\{i : \mu_i \mu_{i+1} < 0 \}$, where
$f = \sum_i\mu_ix^{p_i}$, and no $\mu_i = 0$.
A slight generalization of Descartes' rule of signs \cite{W-04} states that
\begin{equation}
     \#\mathsf{positive\_zeros}(f) \le \#\mathsf{sign\_changes}(f) \label{descartes}
\end{equation}

Consider any $k\times k$ subsquare $M(R,C)$ given by
$R = \{r_1, \ldots, r_k\}, C = \{c_1,\ldots, c_k\}\subset [n]$, and any
nonzero vector $\mu\in\R^k$.  Then $M(R,C) \cdot \mu$ matches
the signomial $f(x) = \sum \mu_ix^{p_{c_i}}$ evaluated at
$x = r_1, \ldots, r_k$.  Using (\ref{descartes}), $\#\{i: f(r_i) = 0\} \le
\#\mathsf{sign\_changes}(f) < k$, so that some $f(r_i)\ne 0$, and
$M(R,C)\mu\ne 0$.  Hence the subsquare has a trivial kernel, and
is nonsingular.
\epf

To avoid problems of precision, we will choose powers of $p_j$ to be consecutive 
integers beginning at $0$. This yields the Vandermonde matrix over $\mathbb{Q}$. It 
can also be shown, by an elementary cardinality argument, that a random matrix 
(using a non-atomic distribution) is unsparsifiable with probability 1 
over infinite 
fields. The above lemma avoids any probability and allows us to 
construct such a matrix as quickly as we can iterate over the entries.


\subsection{Reduction for matrix problems}

After proving the existence of an unsparsifiable matrix in the last
section, we can now prove the hardness of approximation of {\sf matrix
sparsification}. We reduce {\sf min unsatisfy} to {\sf matrix
sparsification}. Given an instance $\langle A,y \rangle$ of {\sf min
unsatisfy}, we create a matrix $M$ such that {\sf matrix
sparsification} on $M$ solves the instance of {\sf min unsatisfy}.

Before describing the reduction, we outline the intuition behind it. We wish
to create a matrix $M$ with many copies of $y$ and some copies of $A$. The
number of copies of $y$ should greatly outnumber the number of copies of $A$.
The desired approximation bounds will be achieved by guaranteeing that $M$ is
composed mostly of zero entries and of copies of $y$. It follows that
minimizing the number of nonzero entries in
the matrix (solving {\sf matrix sparsification}) will reduce to minimizing
the
number of nonzero entries in the copies of $y$ by finding a sparse linear
combination of $y$ with some other dictionary vectors (solving {\sf min
unsatisfy}).

The construction is as follows: Given an instance $\langle A,y \rangle$ of 
{\sf min unsatisfy} (where $A$ is an $m \times n$ matrix, $y \in R^m$, and
$q \ge p$ are free parameters), take an optimally sparse $(p+q) \times p$
matrix $\binom{I_p}{X}$ 
as given by Lemma \ref{lem:const_opt} and
Theorem \ref{unsparsifiable_theorem}
(where $I_p$ is a $p \times p$ identity matrix),
and create matrix $M_l = \binom{I_p}{X} \otimes y = \binom{I_p 
\otimes y}{X \otimes y}$
(of size $(p+q)m \times p$). Further create matrix $I_q \otimes A$ (of 
size $qm
\times qn$), and
take matrix 0 (of size $pm \times qn$) and form matrix $M_r = 
\binom{0}{I_q \otimes A}$
(of size $(p+q)m \times qn$). Append $M_r$ to the right of $M_l$ to create
matrix
$M=M_l|M_r$ of size $(p+q)m \times (p+qn)$.
We can summarize this construction as
$M = \left(\!\begin{array}{cc}I_p \otimes y & 0 \\ X \otimes y & I_q 
\otimes A \\ \end{array}\!\right)
$.

$M_l$ is composed of $p+pq$ $m$-length vectors, all corresponding to 
copies of $y$. $M_r$ is composed of $qn$ $m$-length vectors, all 
corresponding copies of vectors in $A$. By choosing $p=q=n^2$, we ensure 
that the term $pq$ is larger than $qn$ by a factor of $n$. Note that $M$ 
now contains $O(n^3)$ columns.

It follows that the number of zeros in $M$ depends mostly on the number of
zeros induced by a linear combination of dictionary vectors that include
$y$. Because $M_l$ is unsparsifiable, vectors in the rows of $M_l$ will not
contribute to sparsifying other vectors in these rows; only vectors in $M_r$ 
(which are copies of the vectors of $A$) may sparsify vectors in $M_l$ 
(which are copies of the vectors in $y$). It follows that an approximation 
to {\sf matrix sparsification} will yield a similar approximation -- within a 
factor of $1+ n^{-\frac{1}{3}}$ -- to {\sf min unsatisfy}, and that {\sf matrix 
sparsification} is hard to approximate within a factor 
$2^{\log^{.5-o(1)}n^{1/3}} = 2^{\log^{.5-o(1)}n}$ of optimal (assuming NP does not 
admit quasi-polynomial time deterministic algorithms).

\section{Solving {\sf matrix sparsification} through {\sf min unsatisfy}
}\label{relation}

In the previous section we showed that {\sf matrix sparsification} is hard
to approximate. This motivates the search for heuristics and algorithms
under simplifying assumptions for {\sf matrix sparsification}.
In this section we show how to extend algorithms and heuristics for
{\sf min unsatisfy} to apply to {\sf matrix sparsification} -- and hence
{\sf sparse null space} -- while preserving approximation guarantees.
(Note that this result is distinct from the hardness result; neither one
implies the other.)

We first present an algorithm for {\sf matrix sparsification} which is
in essence identical to the one given by Coleman and Pothen \cite{CP-86}
for {\sf sparse null space}. The algorithm assumes the existence of an
oracle for a problem we will call the {\sf sparsest independent vector}
problem. The algorithm makes a polynomial number of queries to this
oracle, and yields an optimal solution to {\sf matrix sparsification}.

The {\sf sparsest independent vector} problem takes full-rank input matrices $A$ and
$B$, where the columns of $B$ are a contiguous set of right-most columns from $A$
(informally, one could say that $B$ is a suffix of $A$, in terms of columns).
The output is the sparsest vector in the span of $A$ but not in the span of $B$.
For convenience, we add an extra output parameter --- a column of $A\setminus B$
which can be replaced by the sparsest independent vector while preserving the
span of $A$. More formally, {\sf sparsest independent vector} is defined as follows. 
(See \S\ref{app:equiv} for the definition of a problem instance.)

\problem{sparsest independent vector (SIV)}{\\
$\mathcal{I}_\mathsf{SIV} =
	\langle A,B \rangle$;  $A$ is an $m\times n$ full rank matrix
	 with $A = (C | B)$ for some non-empty matrix $C$. \\
$S_\mathsf{SIV}(A,B) =
	\{a:a\in\col(A), a \notin\col(B) \}$  \\
$m_\mathsf{SIV}(\langle A,B\rangle, a) = \nnz(a)$
}

The following algorithm reduces {\sf matrix sparsification} on an
$m\times n$ input matrix $A$ to making a polynomial number of queries
to an oracle for {\sf sparsest independent vector}:

\begin{tabbing}
Algorithm Matrix\_Sparsification($A$) \\
$B \leftarrow$ null \\
for \= $i=n$ to $1$:\\
 \> $\langle b_i, a_j\rangle = \mathsf{SIV}(A,B)$ \\
 \> $A \leftarrow (A \setminus \{ a_j \} | b_i )$ \\
 \> $B \leftarrow (b_i | B)$	\\
return $B$
\end{tabbing}

This greedy algorithm sparsifies the matrix $A$ by generating a new matrix
$B$ one column at a time. The first-added column $(b_n)$ is the sparsest possible, and
each subsequent column is the next sparsest. It is decidedly non-obvious
why such
a greedy algorithm would actually succeed; we refer the reader to
\cite{CP-86}
where it is proven that greedy algorithms yield an optimal result on matroids
such as the set of vectors in $\col(A)$. Our first contribution is in
expanding the
result of \cite{CP-86} as follows.

\begin{lemma}
Let subroutine $\mathsf{SIV}$ in algorithm Matrix\_Sparsification be a
$\lambda$-approx\-imation oracle for {\sf sparse independent vector}. Then
the algorithm yields a $\lambda$-approximation to {\sf matrix
sparsification}.
\end{lemma}

\bpf
Given $m\times n$ matrix $A$, suppose $\tilde C$ exactly solves
$\mathsf{MS}(A)$, and that the columns $\tilde c_1, \ldots, \tilde c_n$
of $\tilde C$ are sorted in decreasing order by number of nonzeros.  Let
$s_i = ||\tilde c_i||_0$; then $s_1 \ge s_2 \ge \ldots \ge s_n$. As
already mentioned, given a true oracle to {\sf sparsest independent vector},
algorithm Matrix\_Sparsification would first discover a column with $s_n$
nonzeros, then a column with $s_{n-1}$ nonzeros, etc.

Now suppose algorithm Matrix\_Sparsification made calls to a
$\lambda-$approx\-imation oracle for {\sf sparse independent vector}.
The first column generated by the algorithm, call it $b_n$, will have at most
$\lambda s_n$ nonzeros, since the optimal solution has $s_n$ nonzeros.
The second column generated will have at most $\lambda s_{n-1}$ nonzeros,
since the optimal solution to the call to $\mathsf{SIV}$ has no more than
$s_{n-1}$ nonzeros: even if $b_n$ is suboptimal, it is true that at least one of
$\tilde c_n$ or $\tilde c_{n-1}$ is an optimal solution to
$\mathsf{SIV}(A,b_n)$.

More generally, the $i\up{th}$ column found by the algorithm has
no more then $\lambda s_i$ nonzeros, since at least one of
$\{\tilde c_n,\ldots, \tilde c_i\}$ is an optimal solution to
the $i\up{th}$ query to $\mathsf{SIV}$. Thus we have
$ \nnz(B) = \sum_i ||b_i||_0 \le \sum \lambda ||\tilde c_i||_0
 = \lambda\; \nnz(\tilde C) $, and may conclude that the algorithm yields
a $\lambda-$approximation to {\sf matrix sparsification}.
\epf

It follows that in order to utilize the aforementioned algorithm for
{\sf matrix sparsification}, we need some algorithm for
{\sf sparsest independent vector}. This is in itself problematic, as
the {\sf sparsest independent vector} problem is hard to approximate --
in fact, we will demonstrate later that {\sf sparsest independent vector}
is as hard to approximate as {\sf min unsatisfy}. Hence, although we have
extended the algorithm of \cite{CP-86} to make use of an approximation
oracle for {\sf sparsest independent vector}, the benefit of this
algorithm remains unclear.

To this end, we will show how to solve {\sf sparsest independent vector}
while making queries to an approximate oracle for {\sf min unsatisfy}. This
algorithm preserves the approximation ratio of the oracle. This implies
that {\em
all} algorithms for {\sf min unsatisfy} immediately carry over to
{\sf sparsest independent vector}, and further that they carry over to
{\sf matrix sparsification} as well. This also implies a useful tool
for applying heuristics for {\sf min satisfy} to the other problems.

The problem 
{\sf sparsest independent vector} on input $\langle A,B \rangle$
asks to find the sparsest vector in the span of $A$ but not in the
span of $B$.
It is not difficult to see
that {\sf min unsatisfy} solves a similar problem: Given a matrix $A$ and
target vector $y$ not in the span of $A$,
find the sparsest vector in the span of $(A | y)$ but not in the
span of $A$.
Hence, if we query the oracle for
{\sf min unsatisfy} once for each vector $a_j\notin\col(B)$, one of these
queries
must return the solution for the {\sf sparsest independent vector}
problem. This
discussion implies the following algorithm:

\begin{tabbing}
Algorithm Sparse\_Independent\_Vector($A,B$) \\
$s \leftarrow m+1$		\\
for \= $j=1$ to $n$:	\\
 \> if \= $a_j \notin col(B):$ \\
 \> \> $A_j \leftarrow A \backslash \{a_j\}$ \\
 \> \> $x \leftarrow \mathsf{MU}(A_j, a_j)$ \\
 \> \> $c' \leftarrow A_jx - a_j$ \\
 \> \> if \= $||c'||_0 < s$	\\
 \> \> \> $c \leftarrow c'$;\; 
$s \leftarrow ||c||_0$;\;
$\alpha \leftarrow a_j$	\\
return $\langle c, \alpha \rangle$
\end{tabbing}

Note that when this algorithm is given a $\lambda$-approximate oracle for
{\sf min unsatisfy}, it yields a $\lambda$-approximate algorithm for
{\sf sparsest independent vector}. (In this case, the approximation 
algorithm is valid over the field for which the oracle is valid.)

We conclude this section by giving hardness results for
{\sf sparsest independent vector} by reduction from {\sf min unsatisfy};
we show that any instance
$\langle A,b \rangle$ of {\sf min unsatisfy} may be modeled as an instance
$\langle A',B' \rangle$ of {\sf sparsest independent vector}:
Let $A' = A|y$, and $B' = A$. This suffices to force the linear
combination to
include $y$. It follows that {\sf sparsest independent vector} is as hard to
approximate as {\sf min unsatisfy}, and in fact that the two problems are
approximation equivalent.

\subsection{Approximation algorithms}

We have presented a tool for extending algorithms and heuristics for
{\sf exact dictionary representation} to {\sf min unsatisfy} and then
directly to the matrix problems. When these algorithms make assumptions on
the dictionary of {\sf EDR}, it is necessary to investigate how these
assumptions carry over to the other problems.

To this end, we consider here one of the most popular heuristic for {\sf
EDR} -- $\ell_1$-minimization -- and the case where it is guaranteed to
provide the optimal result. The heuristic is to find a vector $v$ that
satisfies $Dv=s$, while
minimizing $||v||_1$ instead of $||v||_0$. (See \cite{T-05,T-06,D-04} for
more
details.)
In \cite{F-04}, Fuchs shows that under the following relatively simple
condition
$\ell_1$-minimization provides the optimal answer to $\mathsf{EDR}$.

In the following, we write $\sgn(x)$ to indicate $\frac{x}{|x|}$, or zero if
$x=0$. Given a matrix $D$ whose columns are divided into two submatrices
$D_0$ and
$D_1$, we may write $D = (D_0\; D_1)$, even though $D_0$ and $D_1$ may not be
contiguous portions of the full matrix.  (The reader may view this as
permuting
the columns of $D$ before splitting into $D_0$ and $D_1$.)

\begin{theorem}[Fuchs]\label{fuchs_thm}
Suppose that $s = Dv$, and that $||v||_0$ is minimal (so that this
$v$ solves $\mathsf{EDR}(D,s)$).  Split $D = (D_0\; D_1)$
so that $D_0$ contains all the columns in the
support of $v$.  Accordingly, we split the vector $v = \binom{v_0}{\vec
0}$, in which
all coordinates of $v_0$ are nonzero.

If there exists a vector $h$ so that $D_0^Th = \sgn(v_0),$  and
$||D_1^Th||_\infty < 1,$ then $||v||_1 < ||w||_1$ for all vectors $w \ne
v$ with
$Dw=s$.
\end{theorem}

We extend this result
to each of our major problems.

\begin{theorem}\label{fuchs_ext_thm}\quad
\textsf{min unsatisfy.}
Suppose, for a given $A, y$ pair, that $x$
minimizes $||y-Ax||_0$.  Split $y = \binom{y_0}{y_1}$ and
$A = \binom{A_0}{A_1}$ so that $A_1$ is maximal such that $y_1 = A_1x$,
and let $v = y_0 - A_0 x$.  If there is a matrix $u$ with $||u||_\infty <
1$ and
$A_1^T u = - A_0^T\sgn(v)$, then our reduction of $\mathsf{MU}(A,y)$ to an
$\ell_1$
approximation of $\mathsf{EDR}(D,s)$ gives the truly optimal answer.

\textsf{matrix sparsification.}
For a given $m\times n$ matrix $B$,
suppose $C$ minimizes $\nnz(C)$ such that $C=BX$ for invertible $X$.
For any $i\in [n]$, split column $c_i = \binom{c_{i,0}}{\vec 0}$
so that $c_{i,0}$ is completely nonzero, and, respectively,
$B = \binom{B_{i,0}}{B_{i,1}}$,
so that $c_{i,0} = B_{i,0} x_i$.
If, for all $i\in [n]$, there exists vector $u_i$ with $||u_i||_\infty<1$ and
$B_{i,1}^Tu_i = - B_{i,0}^T\sgn(c_{i,0})$,
then our reduction algorithm to an $\ell_1$
approximation of $\mathsf{EDR}$ via {\sf min unsatisfy}
will give a truly optimal answer to this {\sf MS} instance.

\textsf{sparse null space}
For a given matrix $A$ with corank $c$, suppose matrix $V$
solves $\mathsf{SNS}(A)$.  For each $i\in [c]$, split
column $v_i = \binom{v_{i,0}}{\vec 0}$ so that $v_{i,0}$
is completely nonzero and, respectively,
$A = (A_{i,0}\; A_{i,1})$ so that $A_{i,0}v_{i,0} = 0$.
If, for all $i\in [c]$, there exists vector $h_i$ with 
$||A_{i,1}^Th_i||_\infty < 1 \quad\text{and}\quad A_{i,0}^Th_i = \sgn(v_{i,0})$,
then our reduction to an $\ell_1$ approximation of $\mathsf{EDR}$ via
{\sf matrix sparsification} and {\sf min unsatisfy} gives a
truly optimal answer to this {\sf SNS} instance.
\end{theorem}

\bpf
{\sf min unsatisfy.}
As in our reduction from {\sf MU} to {\sf EDR}, we find
matrix $D$ with $DA=0$ and vector $s=Dy$.  Then
\[ \binom{\sgn(v_0)}{u} \in \nll(A^T) = \col(D^T)
\implies \exists h : D^Th = \binom{\sgn(v)}{u}. \]
Splitting $D = (D_0\; D_1)$, we see that
$D_0^Th = \sgn(v)$ and $||D_1^Th||_\infty < 1$, exactly
what is required for theorem \ref{fuchs_thm}, showing
that $\ell_1$ minimization gives the answer $D_0v_0$.
Since $D_0v_0 = (D_0\; D_1)\binom{v}{\vec 0}
= D(y-Ax) = s$, this completes the proof.

{\sf matrix sparsification.}
We write $A\setminus i$ to denote matrix $A$ with
the $i^\text{th}$ column removed.
In our reduction of {\sf MS} to {\sf MU}, we need to solve
instances of {\sf MU} over equations of the form $(B\setminus i)x = b_i$.
According to the {\sf MU} portion of this theorem, it suffices to show that
$(B_{i,1}\setminus i)^Tu_i = -(B_{i,0}\setminus i)^T\sgn(c_{i,0})$.
The condition for this portion of the theorem implies this, since
removing any corresponding rows from a matrix equation of the
form $Ax = By$ still preserves the equality.

{\sf sparse null space.}
As in our reduction from {\sf SNS} to {\sf MS}, we find
a matrix $B$ such that $A$ is a full null matrix for $B$.
For any $i$,
let $u_i = A_{i,1}^Th_i$ so that $A^Th_i = \binom{\sgn(v_{i,0})}{u_i}$.
Then $ \binom{\sgn(v_{i,0})}{u_i} \in \col(A^T) = \nll(B^T)$,
and $B_{i,1}^Tu_i = - B_{i,0}^T\sgn(v_{i,0}),$
which is exactly what is necessary for {\sf matrix sparsification}
to function through $\ell_1$ approximation.
\epf

The following intuitive conditions give insight into which matrices
are amenable to $\ell_1$ approximations.
$A^+$ denotes $(A^TA)^{-1}A^T$, the pseudoinverse
of $A$.

\begin{corollary}
{\sf min unsatisfy.}
Suppose matrix $A = \binom{A_0}{A_1}$ is split by an optimal answer
as in theorem \ref{fuchs_ext_thm}.  If $\row(A_0)\subset \row(A_1)$ and
$||(A_1^T)^+A_0^T||_{1,1} < 1$,
our $\ell_1$ approximation scheme will give a truly optimal answer.
\item {\sf matrix sparsification.}
Suppose matrix $B = \binom{B_{i,0}}{B_{i,1}}$ is split by the columns of
an optimal answer $C=BX$ as in theorem \ref{fuchs_ext_thm}.  If, for any $i$,
$\row(B_{i,0})\subset\row(B_{i,1})$ and
$||(B_{i,1}^T)^+B_{i,0}^T||_{1,1} < 1$,
then our $\ell_1$ approximation will give the optimal answer.

{\sf sparse null space.}
Suppose matrix $A = (A_{i,0}\; A_{i,1})$ is split by the columns of
an optimal answer $V$ with $AV=0$ as in theorem \ref{fuchs_ext_thm}.
If $\col(A_{i,0})\subset\col(A_{i,1})$ and
$||A_{i,0}^+A_{i,1}||_{1,1} < 1$ $\forall i$,
then our $\ell_1$ approximation will give an optimal answer.
\end{corollary}

\section{Acknowledgements}

We thank Daniel Cohen for finding an error in an earlier version of 
this paper.

\bibliographystyle{plain}
\bibliography{sparsebib}

\begin{thebibliography}{10}

\bibitem{AK-95}
E.~Amaldi and V.~Kann.
\newblock The complexity and approximability of finding maximum feasible
  subsystems of linear relations.
\newblock {\em Theoretical computer science}, 147(1--2):181--210, 1995.

\bibitem{ABSS-97}
S.~Arora, L.~Babai, J.~Stern, and Z.~Sweedyk.
\newblock The hardness of approximate optima in lattices, codes and linear
  equations.
\newblock {\em JCSS}, 54(2), 1997.

\bibitem{BK-01}
Piotr Berman and Marek Karpinski.
\newblock Approximating minimum unsatisfiability of linear equations.
\newblock {\em Electronic Colloquium on Computational Complexity (ECCC)},
  8(25), 2001.

\bibitem{BH-85}
M.~Berry, M.~Heath, I.~Kaneko, M.~Lawo, R.~Plemmons, and R.~Ward.
\newblock An algorithm to compute a sparse basis of the null space.
\newblock {\em Numer. Math.}, 47:483--504, 1985.

\bibitem{BFP-95}
R.A. Brualdi, S.~Friedland, and A.~Pothen.
\newblock The sparse basis problem and multilinear algebra.
\newblock {\em SIAM Journal on Matrix Analysis and Applications}, 16(1):1--20,
  1995.

\bibitem{CRT-06}
E.~Cand\`es, J.~Romberg, and T.~Tao.
\newblock Robust uncertainty principles: Exact signal reconstruction from
  highly incomplete frequency information.
\newblock {\em IEEE Trans. Inform. Theory}, 52:489--509, 2006.

\bibitem{CM-92}
S.~Frank Chang and S.~Thomas McCormick.
\newblock A hierarchical algorithm for making sparse matrices sparser.
\newblock {\em Mathematical Programming}, 56(1--3):1--30, August 1992.

\bibitem{CW-92}
R.~Coifman and M.~Wickerhauser.
\newblock Entropy-based algorithms for best basis selection.
\newblock {\em IEEE Transactions on Information Theory}, 38(2):713--718, 1992.

\bibitem{CP-86}
T.F. Coleman and A.~Pothen.
\newblock The null space problem {I}. complexity.
\newblock {\em SIAM Journal on Algebraic and Discrete Methods}, 7(4):527--537,
  October 1986.

\bibitem{DKRS-03}
Irit Dinur, Guy Kindler, Ran Raz, and Shmuel Safra.
\newblock Approximating {CVP} to within almost-polynomial factors is {NP}-hard.
\newblock {\em Combinatorica}, 23(2):205--243, 2003.

\bibitem{D-04}
David~L. Donoho.
\newblock For most large underdetermined systems of linear equations, the
  minimal l1-norm solution is also the sparsest.
\newblock http://www-stat.stanford.edu/\~\
  donoho/Reports/2004/l1l0EquivCorrected.pdf, 2004.

\bibitem{DER-86}
I.S. Duff, A.M. Erisman, and J.K. Reid.
\newblock {\em Direct Methods for Sparse Matrices}.
\newblock Oxford University Press, 1986.

\bibitem{F-04}
J.J. Fuchs.
\newblock On sparse representations in arbitrary redundant bases.
\newblock {\em IEEE Trans. Inf. Th.}, 50(6):1341--1344, June 2004.

\bibitem{GGIMS-02}
A.C. Gilbert, S.~Guha, P.~Indyk, S.~Muthukrishnan, and M.~Strauss.
\newblock Near-optimal sparse fourier representations via sampling.
\newblock In {\em STOC}, 2002.

\bibitem{GMS-04}
A.C. Gilbert, S.~Muthukrishnan, and M.~Strauss.
\newblock Improved time bounds for near-optimal sparse fourier representations.
\newblock In {\em Proc. SPIE Wavelets XI}, 2005.

\bibitem{GH-87}
J.R. Gilbert and M.T. Heath.
\newblock Computing a sparse basis for the null space.
\newblock {\em SIAM Journal on Algebraic and Discrete Methods}, 8(3):446--459,
  July 1987.

\bibitem{H-84}
A.J. Hoffman and S.T. McCormick.
\newblock A fast algorithm that makes matrices optimally sparse.
\newblock In {\em Progress in Combinatorial Optimization}. Academic Press,
  1984.

\bibitem{JP-78}
D.S. Johnson and F.P. Preparata.
\newblock The densent hemisphere problem.
\newblock {\em Theoret. Comput. Sci.}, 6:93--107, 1978.

\bibitem{K-94}
Viggo Kann.
\newblock Polynomially bounded minimization problems that are hard to
  approximate.
\newblock {\em Nordic Journal of Computing}, 1(3):317--331, Fall 1994.

\bibitem{KMMP-04}
T.~Kavitha, K.~Mehlhorn, D.~Michail, and K.~Paluch.
\newblock A faster algorithm for minimum cycle basis of graphs.
\newblock In {\em Automata, Languages, and Programming: 31st International
  Colloquium, ICALP 2004 Proceedings}, Lecture Notes in Computer Science.
  Springer, 2004.

\bibitem{MG-07}
J.~Matou\u{s}ek and B.~Gartner.
\newblock {\em Understanding and Using Linear Programming}.
\newblock Springer, 2007.

\bibitem{M-83}
S.T. McCormick.
\newblock {\em A combinatorial approach to some sparse matrix problems}.
\newblock PhD thesis, Stanford Univ., Stanford, California, 1983.
\newblock Technical Report 83-5, Stanford Optimization Lab.

\bibitem{M-04}
S.~Muthukrishnan.
\newblock Nonuniform sparse approximation with haar wavelet basis.
\newblock DIMACS TR:2004-42, 2004.

\bibitem{N-95}
B.K. Natarajan.
\newblock Sparse approximate solutions to linear systems.
\newblock {\em SIAM J. on Comput.}, 24(2):227--234, 1995.

\bibitem{NT-09}
D.~Needell and J.~Tropp.
\newblock Cosamp: Iterative signal recovery from incomplete and inaccurate
  samples.
\newblock {\em Applied and Computational Harmonic Analysis}, 26(3):301--322,
  2009.

\bibitem{N-06}
T.~Neylon.
\newblock {\em Sparse solutions for linear prediction problems}.
\newblock PhD thesis, New York University, New York, New York, 2006.

\bibitem{P-84}
A.~Pothen.
\newblock {\em Sparse null bases and marriage theorems}.
\newblock PhD thesis, Cornell University, Ithica, New York, 1984.

\bibitem{S-07}
E.~Schmidt.
\newblock Zur thoerie der linearen und nichtlinearen integralgleichungen, i.
\newblock {\em Math. Annalen.}, 64(4):433--476, 1906--1907.

\bibitem{S-89}
S.~Singhal.
\newblock Amplitude optimization and pitch predictin in multi-pulse coders.
\newblock {\em IEEE Transactions on Acoustics, Speech and Signal Processing},
  37(3), March 1989.

\bibitem{SS-00}
A.~Smola and B.~Sch{\"o}lkopf.
\newblock Sparse greedy matrix approximation for machine learning.
\newblock In {\em Proc. 17th International Conf. on Machine Learning}, pages
  911--918. Morgan Kaufmann, San Francisco, CA, 2000.

\bibitem{T-03}
V.~Temlyakov.
\newblock Nonlinear methods of approximation.
\newblock {\em Foundations of Comp. Math.}, 3(1):33--107, 2003.

\bibitem{T-04}
J.A. Tropp.
\newblock Greed is good: algorithmic results for sparse approximation.
\newblock {\em IEEE Transactions on Information Theory}, 50(10):2231--2242,
  October 2004.

\bibitem{T-06}
J.A. Tropp.
\newblock Just relax: convex programming methods for subset selection and
  sparse approximation.
\newblock {\em IEEE Transactions on Information Theory}, 50(3):1030--1051,
  March 2006.

\bibitem{T-05}
J.A. Tropp.
\newblock Recovery of short, complex linear combinations via $\ell_1$
  minimization.
\newblock {\em IEEE Trans. Inform. Theory}, 51(1):188--209, January 2006.

\bibitem{W-04}
X.~Wang.
\newblock A simple proof of descartes' rule of signs.
\newblock {\em Amer. Math. Monthly}, 111:525--526, 2004.

\bibitem{ZZ-05}
X.~Zhao, X.~Zhang, T.~Neylon, and D.~Shasha.
\newblock Incremental methods for simple problems in time series: algorithms
  and experiments.
\newblock In {\em Ninth International Database Engineering and Applications
  Symposium (IDEAS 2005)}, pages 3--14, 2005.

\bibitem{ZGSD-06}
J.~Zhou, A.~Gilbert, M.~Strauss, and I.~Daubechies.
\newblock Theoretical and experimental analysis of a randomized algorithm for
  sparse fourier transform analysis.
\newblock {\em Journal of Computational physics}, 211:572--595, 2006.

\end{thebibliography}


\appendix

\section{Approximation equivalence}\label{app:equiv}

Here we define approximation equivalence. Some of our notation and
definitions are inspired by \cite{AK-95}, which itself built upon
\cite{K-94}.

\begin{definition}
An {\em optimization problem} is a four-tuple
$F = \{\mathcal{I}_F, S_F, M_F,\opt_F\}$, where
$\mathcal{I}_F$ is the set of input instances,
$S_F(x)$ is the solution space for $x \in \mathcal{I}_F$,
$M_F(x,y)$ is the objective metric for $x \in \mathcal{I}_F$ and $y \in
S_F(x)$, and
$\opt_F \in \{$min,max$\}$.
\end{definition}

We will assume throughout the paper that $\opt_F=$ min.

For any optimization problem $F$ and $x \in \mathcal{I}_F$, we define
$F(x) = \argmin_{y \in S_F(x)}M_F(x,y)$ and $||F(x)||=M_F(x,F(x))$. An
approximation $\tilde{F}$ to $F$
is any map on $\mathcal{I}_F$ with $\tilde{F}(x) \in S_F(x)$. We write
$||\tilde{F}(x)||$ for $M_F(x,\tilde{F}(x))$. $\tilde{F}$ is an
$\lambda$-{\em approximation} for $F$ when, for all $x \in \mathcal{I}_F$,
$\frac{||\tilde{F}(x)||}{||F(x)||} \le \lambda(|x|)$.

\begin{definition}\label{equiv_def}
Given optimization problems $F$ and $G$, an {\em exact reduction} from
$F$ to $G$ is a pair $\langle t_1,t_2 \rangle$ that satisfies the
following:
(1) $t_1,t_2 \in P$.
(2) $t_1: \mathcal{I}_F \rightarrow \mathcal{I}_G$ and for all
	$x \in \mathcal{I}_F$,  $y\in S_G(t_1(x))$, we have $t_2(x,y) \in S_F(x)$.
(3) For all $x \in \mathcal{I}_F$, $y\in S_G(t_1(x))$, we have
	$M_F(x,t_2(x,y)) = M_G(t_1(x),y)$.
(4) For all $x \in \mathcal{I}_F$, $||F(x)|| \ge ||G(t_1(x))||$.

We write $F \preceq G$.
We write $F\sim G$ to denote that $F\preceq G$ and $G\preceq F$, and
call these problems {\em equivalent}.
\end{definition}

\begin{theorem}
If $F \preceq G$ and $G$ admits a $\lambda$-approximation, then so does
$F$.
\end{theorem}

\bpf
We are given that for $G$, there exists $\tilde{G}$ with
$\frac{||\tilde{G}(x)||}{||G(x)||} \le \lambda$ for all $x \in
\mathcal{I}_G$.
Let $\tilde{F}(x) = t_2(x,\tilde{G}(t_1(x)))$, where
$\langle t_1,t_2 \rangle$ is the $F \preceq G$ exact reduction. It
suffices to show that
$\frac{||\tilde{F}(x)||}{||F(x)||} \le
\frac{||\tilde{G}(x')||}{||G(x')||} $,
where $x' = t_1(x)$. By the fourth
item of the definition, it suffices to demonstrate that
$||\tilde{F}(x)|| \le ||\tilde{G}(x')||$. By the third item,
$||\tilde{F}(x)|| = M_F(x,t_2(x,\tilde{G}(x'))) =
M_G(x',\tilde{G}(x')) = ||\tilde{G}(x')||$.
\epf

In fact, it can be shown that
$\frac{||\tilde{F}(x)||}{||F(x)||} =
\frac{||\tilde{G}(x')||}{||G(x')||} $.

\begin{corollary}\label{equiv_cor}
If $F \sim G$, then $F$ admits a $\lambda$-approximation if and only if
$G$ admits a $\lambda$-approximation.
\end{corollary}

\section{Relaxed Versions}\label{relaxed_section}

The following problems are variations which work with
more approximate solution spaces.  This can be considered as allowing some
noise in either the inputs or outputs.  It is not difficult to extend the above proofs
to see that these four problems are equivalent as well.

\problem{Relaxed Dictionary Representation (RDR)}{\\
$\mathcal{I}_\mathsf{RDR} = \langle D,s,\delta\rangle$,
$m\times n$ matrix $D$, vector $s$ with $s\in \col(D)$, $\delta \ge 0$\\
$S_\mathsf{RDR}(D,s,\delta) = \{\langle v,w\rangle$ each in $\R^n : D(v-w) = s, ||w||\le\delta \}$ \\
$m_\mathsf{RDR}(\langle D,s\rangle, \langle v,w\rangle) = ||v||_0$
}

\problem{Relaxed MinUnsatisfy (RMU)}{\\
$\mathcal{I}_\mathsf{RMU} = \langle A,y, \delta\rangle$,
$m\times n$ matrix $A$, vector $y\in \R^m$, $\delta\ge 0$\\
$S_\mathsf{RMU}(A,y,\delta) = \{\langle x\in \R^n, w\in \R^m\rangle: ||w|| \le \delta\}$ \\
$m_\mathsf{RMU}(\langle A,y\rangle, \langle x,w\rangle) = ||y-Ax+w||_0$
}

\problem{Relaxed Sparse Null Space (RSNS)}{\\
$\mathcal{I}_\mathsf{RSNS} = \langle A,\delta\rangle$,
matrix $A$, and $\delta \ge 0$ \\
$S_\mathsf{RSNS}(A,\delta) = \{\langle M,N\rangle: N $
is a full null matrix for $A, ||M|| \le \delta\}$\\
$m_\mathsf{RSNS}(\langle A,\delta\rangle, \langle M,N\rangle) = \nnz(M+N)$
}

\problem{Relaxed Matrix Sparsification (RMS)}{\\
$\mathcal{I}_\mathsf{RMS} = \langle B, \delta\rangle$,
matrix $B$, and $\delta\ge 0$ \\
$S_\mathsf{RMS}(B,\delta) = \{\langle M,N\rangle: N = BX, X $
invertible, $||M|| \le \delta\}$ \\
$m_\mathsf{RMS}(\langle B,\delta\rangle, \langle M,N\rangle) =
\nnz(M+N)$
}

\end{document}